\input harvmac
\input epsf

\newcount\figno
\figno=0
\def\fig#1#2#3{
\par\begingroup\parindent=0pt\leftskip=1cm\rightskip=1cm\parindent=0pt
\baselineskip=11pt
\global\advance\figno by 1
\midinsert
\epsfxsize=#3
\centerline{\epsfbox{#2}}
\vskip 12pt
{\bf Fig. \the\figno:} #1\par
\endinsert\endgroup\par
}
\def\figlabel#1{\xdef#1{\the\figno}}
\def\encadremath#1{\vbox{\hrule\hbox{\vrule\kern8pt\vbox{\kern8pt
\hbox{$\displaystyle #1$}\kern8pt}
\kern8pt\vrule}\hrule}}

\overfullrule=0pt

\noblackbox
\parskip=1.5mm

\def\Title#1#2{\rightline{#1}\ifx\answ\bigans\nopagenumbers\pageno0
\vskip0.5in
\else\pageno1\vskip.5in\fi \centerline{\titlefont #2}\vskip .3in}


\noblackbox
\parskip=1.5mm

  
\def\npb#1#2#3{{\it Nucl. Phys.} {\bf B#1} (#2) #3 }
\def\plb#1#2#3{{\it Phys. Lett.} {\bf B#1} (#2) #3 }
\def\prd#1#2#3{{\it Phys. Rev. } {\bf D#1} (#2) #3 }
\def\prl#1#2#3{{\it Phys. Rev. Lett.} {\bf #1} (#2) #3 }
\def\mpla#1#2#3{{\it Mod. Phys. Lett.} {\bf A#1} (#2) #3 }
\def\ijmpa#1#2#3{{\it Int. J. Mod. Phys.} {\bf A#1} (#2) #3 }

\def\cmp#1#2#3{{\it Commun. Math. Phys.} {\bf #1} (#2) #3 }

\def\bb#1{{\tt hep-th/#1}}

\def\heph#1{{\tt hep-ph/#1}}

\def\rmp#1#2#3{{\it Rev. Mod. Phys.} {\bf #1} (#2) #3 }
\def\jetplet#1#2#3{{\it JETP Letters} {\bf #1} (#2) #3}

\def\ma#1#2#3{{\it Math. Ann.  } {\bf #1} (#2) #3 }
\def\npps#1#2#3{{\it Nucl. Phys. Proc. Suppl. } {\bf #1} (#2) #3 }
\def\atmp#1#2#3{{\it Adv. Theor. Math. Phys.} {\bf #1} (#2) #3 }
\def\jhep#1#2#3{{\it JHEP} {\bf #1} (#2) #3}


           \def\CO{{\cal O}} 
  \def\CF{{\cal F}} 
\def\CL{{\cal L}}   
  \def\CD{{\cal D}} 
  \def\CE{{\cal E}}
\def\CN{{\cal N}} \def\CS{{\cal S}}

\def\sumint{\hbox{$\sum$}\!\!\!\!\!\!\int}

\def\simsub{\mathop{\sim}}


\def\dj{\hbox{d\kern-0.347em \vrule width 0.3em height 1.252ex depth
-1.21ex \kern 0.051em}}

\def\half{{1\over 2}\,}


\lref\tf{A. Fotopoulos and T.R. Taylor, \prd{59}{1999}{061701} (\bb{9811224}).}
\lref\az{P. Arnold and C. Zhai, \prd{50}{1994}{7603} (\heph{9408276}); 
\prd{51}{1995}{1906}
(\heph{9410360}).}
\lref\os{M.A.R. Osorio, \ijmpa{7}{1992}{4275.}}
\lref\witsusy{E. Witten, \npb{202}{1982}{253.}}
\lref\rbs{W. Siegel, \plb{84}{1979}{193\semi}
D.M. Caper, D.R.T. Jones and P. van Nieuwenhuizen, \npb{167}{1980}{479.}}
\lref\rbvm{J.L.F. Barb\'on and M.A. V\'azquez-Mozo, \npb{475}{1996}{244} 
(\bb{9605050}).}
\lref\repstein{P. Epstein, \ma{56}{1903}{615}; \ma{63}{1907}{205.}}
\lref\rdhoker{E. D'Hoker, \npb{201}{1982}{401.}}
\lref\rmt{R. Dijkgraaf, E. Verlinde and H. Verlinde, \npb{500}{1997}{43} 
(\bb{9703030}); \npps{62}{1988}{348} (\bb{9709107}).} 
\lref\rkap{J.I. Kapusta, {\it Thermal Field Theory}, Cambridge 1989.}
\lref\shenker{S.H. Shenker, {\it Another length scale in string theory?},
(\bb{9509132}).}
\lref\rbvmd{J.L.F. Barb\'on and M.A. V\'azquez-Mozo, \npb{497}{1997}{236} 
(\bb{9701142}).}
\lref\rbkr{J.L.F. Barb\'on and E. Rabinovici, \npb{545}{1999}{371} (\bb{9805143})\semi
J.L.F. Barb\'on, I.I. Kogan and E. Rabinovici, \npb{544}{1999}{104}
(\bb{9809033}).}
\lref\rabkr{S.A. Abel, J.L.F. Barb\'on, I.I. Kogan and E. Rabinovici, \jhep{04}{1999}{015} 
(\bb{9902058}).}
\lref\rhp{G.T. Horowitz and J. Polchinski, \prd{55}{1997}{6189} (\bb{9612146});
\prd{57}{1998}{2557} (\bb{9707170}).}
\lref\rsathi{S.K. Rama and B. Sathiapalan, \mpla{13}{1998}{3147} 
(\bb{9810069})\semi
S. Bal and B. Sathiapalan, {\it High temperature limit of the $N=2$
matrix model}, (\bb{9902087}).}
\lref\rmlv{M. Li, E. Martinec and V. Sahakian, \prd{59}{1999}{044035} 
(\bb{9809061})\semi
E. Martinec and V. Sahakian,
\prd{59}{1999}{124005} (\bb{9810224});
{\it Black holes and five-brane thermodynamics}, (\bb{9901135})}
\lref\rtherm{S. Frautschi, \prd{3}{1971}{2821\semi}
R.D. Carlitz, \prd{5}{1972}{3231\semi}
N. Cabbibo and G. Parisi, \plb{59}{1975}{67\semi}
E. Alvarez, \prd{31}{1985}{418;} \npb{269}{1986}{596\semi}
M.J. Bowick and L.C.R. Wijewardhana, \prl{54}{1985}{2485\semi}
J. Polchinski, \cmp{104}{1986}{37\semi}
Ya.I. Kogan, \jetplet{45}{1987}{709\semi}
E. Alvarez and M.A.R. Osorio, \prd{36}{1987}{1175;} \npb{304}{1988}{327\semi}
J.J. Atick and E. Witten, \npb{310}{1988}{291\semi}
R. Brandenberger and C. Vafa, \npb{316}{1989}{391.}
}
\lref\rmicro{N. Deo, S. Jain and C.-I. Tan, \plb{220}{1989}{125;} 
\prd{40}{1989}{2646.}}
\lref\bh{J.M. Maldacena, {\it Black Holes in String Theory}, Princeton Ph.D.
Thesis (\bb{9607235}).}
\lref\rmalda{J.M. Maldacena, \atmp{2}{1998}{231} (\bb{9711200}).}
\lref\rwads{E. Witten, \atmp{2}{1998}{253} (\bb{9802150}); \atmp{2}{1998}{505} 
(\bb{9803131})\semi
S.S. Gubser, I.R. Klebanov and A.M. Polyakov, \plb{428}{1998}{105} (\bb{9802109}).} 
\lref\rdmvv{R. Dijkgraaf, G. Moore, E. Verlinde and H. Verlinde, \cmp{185}{1997}{197}
(\bb{9608096}).}
\lref\rbss{L. Brink, J.H. Schwarz and J. Scherk, \npb{121}{1977}{77.}}
\lref\rpdb{J. Polchinski, \prl{75}{1995}{4724} (\bb{9510017});
\rmp{68}{1996}{1245} (\bb{9607050}).}
\lref\rdbt{S.S. Gubser, I.R. Klebanov and A.W. Peet, \prd{54}{1996}{3915} 
(\bb{9602135})\semi
S.S. Gubser, {\it Thermodynamics of spinning D3-brane}, (\bb{9810225})}
\lref\rkapnp{J.I. Kapusta, \npb{148}{1979}{461\semi}
D.J. Gross, R.D. Pisarski and L.G. Yaffe, \rmp{53}{1981}{43.}}
\lref\rgkt{S.S. Gubser, I.R. Klebanov and A.A. Tseytlin, \npb{534}{1998}{202}
(\bb{9805156})\semi
K. Landsteiner, \mpla{14}{1999}{379} (\bb{9901143})\semi
M.M. Caldarelli and D. Klemm, {\it M-theory and stringy corrections to  
anti-de Sitter black holes and conformal field theories}, (\bb{9903078}).}
\lref\rmwi{N. Itzhaki, J.M. Maldacena, J. Schonnenschein and S. Yankielowicz,
\prd{58}{1998}{046004} (\bb{9802042}).}
\lref\riibm{N. Ishibashi, H. Kawai, Y. Kitazawa and A. Tsuchiya, \npb{498}{1997}{467}
(\bb{9612115}).}
\lref\rhm{D.A. Lowe, \plb{403}{1997}{243} (\bb{9704041}).}
\lref\rgrisem{G. Grignani and G.W. Semenoff, {\it Thermodynamics partition function
of matrix superstrings}, (\bb{9903246})\semi
J.P. Pe\~nalba, {\it Non-perturbative thermodynamics in Matrix string theory},~
(\bb{9904094}).}
\lref\rmodft{M.B. Green, \npb{381}{1992}{201\semi}
M.A. V\'azquez-Mozo, \plb{388}{1996}{494} (\bb{9607052})\semi
S. Lee and L. Thorlacius, \plb{413}{1997}{303} (\bb{9707167})\semi
M.L. Meana, M.A.R. Osorio and J.P. Pe\~nalba, \npb{531}{1998}{613} (\bb{9803058})\semi
S. Chaudhuri and D. Minic, \plb{433}{1998}{301} (\bb{9803120})\semi
J. Ambjorn, Yu.M. Makeenko and G.W. Semenoff, \plb{445}{1999}{307} (\bb{9810170})\semi
M.L. Meana and J.P. Pe\~nalba, \plb{447}{1999}{59} (\bb{9811170}).}
\lref\rml{M. Li, \jhep{03}{1999}{004} (\bb{9807196}).}
\lref\rbraaten{E. Braaten, \prl{74}{1995}{2164} (\heph{9409434}).}
\lref\rr{C. Kim and S.-J. Rey, {\it Thermodynamics of large-N super
Yang-Mills theory and AdS/CFT correspondence}, (\bb{9905205}).}
\lref\rty{A.A. Tseytlin and S. Yankielowicz, \npb{541}{1999}{145} (\bb{9809032}).}
\lref\rfive{C. Zhai and B. Kastening, \prd{52}{1995}{7232} (\heph{9507380}).}
\lref\rfived{E. Braaten and A. Nieto, \prd{53}{1996}{3421} (\heph{9510408}).}


\line{\hfill ITFA-99-07}  
\line{\hfill SPIN-1999/09}
\line{\hfill {\tt hep-th/9905030}}

\Title{\vbox{\baselineskip 12pt\hbox{}
 }}
{\vbox{{\centerline{A note on Supersymmetric Yang-Mills thermodynamics}}
}}


\centerline{$\quad$ {M.A. V\'azquez-Mozo\foot{vazquez@wins.uva.nl, 
M.Vazquez-Mozo@phys.uu.nl}
 }}

\medskip

\centerline{{\sl Instituut voor Theoretische Fysica}}
\centerline{{\sl Universiteit van Amsterdam}}
\centerline{{\sl Valckenierstraat 65}}
\centerline{{\sl 1018 XE Amsterdam, The Netherlands}}

\centerline{{and}}

\centerline{{\sl Spinoza Instituut}}
\centerline{{\sl Universiteit Utrecht}}
\centerline{{\sl Leuvenlaan 4}}
\centerline{{\sl 3584 CE Utrecht, The Netherlands}}

 \vskip 1.2cm

\noindent
The thermodynamics of supersymmetric Yang-Mills theories is studied by computing
the two-loop correction to the canonical free energy and to the equation of state 
for theories with 16, 8 and 4 supercharges in any dimension $4\leq d\leq 10$, and in
two dimensions at finite volume. In the four-dimensional case we also evaluate the 
first non-analytic contribution in the 't Hooft coupling to the free energy, 
arising from the resummation of ring diagrams. To conclude, we discuss some 
applications to the study of the Hagedorn transition in string theory in the 
context of Matrix strings and speculate on the possible
physical meaning of the transition.


\Date{5/99}


\newsec{Introduction}

The study of the thermal properties of supersymmetric Yang-Mills (SYM) and 
superstring theories has received a boost \refs\rmodft\refs\rbvmd\refs\rsathi
\refs\rbkr\refs\rabkr\refs\rgrisem\ after the D-brane revolution 
\refs\rpdb. More recently, the conjecture by
Maldacena of a duality between string theory on anti-de Sitter (AdS) 
backgrounds and the
large-N limit of SYM theories \refs\rmalda\refs\rmwi\ has further motivated the
study of these issues in the hope that it could lead to a clarification of
the mechanisms of confinement in non-abelian gauge theories 
\refs\rwads.

Beyond its purely cosmological/phenomenological relevance, the study of SYM
thermodynamics finds interesting applications in the study of near-extremal black
holes \refs\bh\ and D-branes \refs\rdbt\refs\rmlv. When looking at the effective 
theory of 
coincident D-branes there are, generically, two regimes
associated with the values of the effective gauge couplings $g_{s}Q$
($Q$ being the number of coincident D-branes or the charges of the black hole).
When $g_{s}Q<1$ the field theory limit is well described by a perturbative
SYM theory living on the world-volume of the D-brane. On the other hand, taking 
$g_{s}Q>1$ the weakly coupled D-brane picture is not appropriate any more.
Using Maldacena's conjecture, however, it is possible to relate the field
theory limit in this non-perturbative regime (large 't Hooft coupling) with 
a supergravity computation on some background of the form $AdS_{d}\times ({\rm 
Spheres})$. Corrections to the leading strong coupling result of order 
$\CO(1/g_{s}Q)$ are then associated with higher dimensional terms ($\alpha'$
corrections) in the supergravity effective action.

The study of the corrections to the leading results on both sides (weak \refs\tf\ 
and strong \refs\rgkt\refs\rty\ coupling) has been done for the conformal 
$\CN=4$ SYM theory in $d=4$
and a tendency in both curves to meet was detected (see however \refs\rml). 
In this note we will try to 
achieve a twofold objective. First, to compute the two-loop free energy of 
SYM theories with 16, 8 and 4 supercharges in various dimensions. 
The first class of theories are specially interesting because of their 
potential application to non-conformal versions of the Maldacena conjecture 
\refs\rmwi. Second, to 
obtain the first non-analytic correction in
the 't Hooft coupling, of order $\CO[(g_{YM}N)^{3/2}]$,
to the two-loop result in four dimensional theories arising from the resummation of 
ring diagrams. We will analyze also with
some detail the two-dimensional case, where the infrared divergences will
be handled by putting the system at finite volume.

The physics of the high temperature string gas has been a recurrent
issue in string theory (for a sample of papers from the ``golden age'' 
of thermal strings see \refs\rtherm\refs\rmicro). In recent 
years we have gained interesting insights about the physical meaning 
of the Hagedorn divergence, in spite of the fact that a full and 
detailed understanding of the problem seems to be still at large.  
Although this will not be the main subject of this note, we will 
try to discuss some aspects of the Hagedorn transition that could 
be enlightened by our results on SYM thermodynamics. In particular we will
use our study of the two-dimensional SYM theories to try to get some
qualitative information about the Hagedorn transition using Matrix 
strings as a non-perturbative definition of Type-IIA superstrings. 
In spite of being non-conclusive, we hope the discussion to be helpful
in shedding some light to such a confusing issue.

The present paper is organized as follows: in the next Section,
the two-loop corrections to the thermal free
energy of SYM theories with 16, 8 and 4 supercharges 
will be computed in any dimension $4\leq d\leq 10$ using dimensional 
reduction from the corresponding maximal 
$\CN=1$ SYM theory. In Section 2.3 we compute the next correction to the two-loop
free energy for SYM theories in four dimensions. Section 2.4 will be devoted
to the study of the two-dimensional case at finite volume.
Finally, in Section 3 we will summarize the conclusions and discuss some possible 
application of our results to the study of the Hagedorn transition in Matrix
string theory.

\newsec{Two-loop free energy of SYM theories}

\subsec{Supersymmetric Yang-Mills theories in various dimensions}

In this section we will compute the next-to-leading contribution to the canonical 
free energy of supersymmetric Yang-Mills theories with 16, 8 and 4 supercharges. 
In order to keep the analysis
general, we will start with ${\cal N}=1$ SYM in $\CD$ dimensions, whose dynamics 
is governed by the
action \refs\rbss\
$$
S=\int d^{\CD}x \, 
{\rm Tr}\left[-{1\over 4g_{YM}^{2}}F_{AB}F^{AB}+i\bar{\psi}\Gamma^{A}D_{A}\psi\right]
$$
where $A,B=0,\ldots,\CD-1$ and both the gauge fields and spinors are in the adjoint 
representation
of $U(N)$. We get theories with different number of supercharges by choosing the 
appropriate
value of ${\CD}$ for which that number $\#_{SC}$ is maximal:
$$
\eqalign{
\#_{SC} = 16 &\rightarrow  {\cal D}_{\rm max}=10  \cr
\#_{SC} =8 &\rightarrow {\cal D}_{\rm max}=6 \cr
\#_{SC} =4 &\rightarrow {\cal D}_{\rm max}=4. }
$$
In addition, different conditions on fermions have to be imposed in order to keep the
number of bosonic and fermionic degrees of freedom equal. Thus, when $\CD_{\rm max}=10$
fermions have to be taken Majorana-Weyl, while for $\CD_{\rm max}=6$ and 
$\CD_{\rm max}=4$ they satisfy Weyl conditions (actually in $d=4$ we can choose the
fermions to be either Majorana or Weyl, both conditions being equivalent \refs\rbss). 
This ensures that the number of physical
bosonic and fermionic degrees of freedom will be equal to $\CD_{\rm max}-2$. 

In general, however, we will be interested in SYM theories with $\#_{SC}$ supercharges 
in dimensions $d\leq {\cal D}_{\rm max}$. This theories can be obtained by
dimensional 
reduction of the corresponding
maximal ${\cal N}=1$ SYM theory in $\CD={\cal D}_{\rm max}$ \refs\rbss. Thus, 
we can parameterize any 
$d$-dimensional
SYM theory with any number of supercharges by specifying both $d$ and the maximal 
dimension 
${\cal D}_{\rm max}$ from which it is obtained by dimensional reduction.
In this way, starting with $\CD_{\rm max}=10$ ($\CN=1$ in $d=10$) we get $\CN=1$ in 
$d=8$, $\CN=2$ in $d=6$, $\CN=4$ in $d=4$ and $\CN=8$ in $d=2$. Starting instead with 
$\CD_{\rm max}=6$ 
($\CN=1$ in $d=6$) we will have $\CN=2$ in $d=4$ and $\CN=4$ in $d=2$. Finally, if 
we take $\CD_{\rm max}
=4$ we can retrieve $\CN=2$ in $d=2$. (for odd dimensions $2n-1$ we have the $\CN$
corresponding to dimension $2n$).

\subsec{Two-loop free energy for $d\geq 4$}

With this in mind, we can proceed to compute the canonical free energy in 
perturbation theory
for any supersymmetric Yang-Mills theory characterized by $({\cal D}_{\rm max},d)$, 
by writing 
down the contribution of vacuum Feynman diagrams of  ${\cal N}=1$ in 
$\CD={\cal D}_{\rm max}$ SYM
and restricting internal momentum in loops to $d$ dimensions. That way 
we are able to keep track
of the contribution of gauge bosons and scalars (as well as their supersymmetric 
partners) 
without having to consider a larger number of diagrams\foot{For example, using 
this trick one
can get the result of ref. \refs\tf\ by computing, instead of ten, only four 
two-loop Feynman 
diagrams.}. The final result, of course, will depend on $(d,{\cal D}_{\rm max})$.

As a warmup exercise, we will compute the one-loop free energy density. In the maximal 
$\CN=1$ theory
we have three diagrams, a bosonic loop, a fermionic loop and the ghost loop, 
which after multiplying
by their corresponding degeneracy factor respectively give (we use the notation of 
ref. \refs\az)
$$
\eqalign{
{\cal F}(\beta)_{\rm 1-loop} & =
{1\over 2}{N}^{2}{\cal D}_{\rm max} \sumint_{(P)}[d^{d}P] \log P^{2}-{1\over 2}
{N}^2({\cal D}_{\rm max}-2)\sumint_{\{P\}}[d^{d}P]\log P^{2}\cr & -
{N}^{2}\sumint_{(P)}[d^{d}P]\log P^{2}}
$$
where by $(P)$ and $\{P\}$ we represent, respectively, bosonic (periodic) and 
fermionic (anti-periodic)
boundary conditions along the Euclidean time and the factor of ${N}^{2}$ is 
due to the fact that 
all fields are in the adjoint representation of $U({N})$. After a straightforward 
computation 
we arrive at
$$
\eqalign{\sumint_{(P)}[d^{d}P]\log P^{2} & \equiv 
{1\over\beta}\sum_{(\omega_{n})}\int {d^{d-1}p\over
(2\pi)^{d-1}} \log (p^{2}+\omega_{n}^{2})= \Lambda_{0}-
{2\Gamma(d/2)\over \pi^{d\over 2}}\zeta(d)\beta^{-d}
\cr 
\sumint_{\{P\}}[d^{d}P]\log P^{2} & \equiv {1\over \beta}\sum_{[\omega_{n}]}
\int {d^{d-1}p\over
(2\pi)^{d-1}} \log (p^{2}+\omega_{n}^{2})= \Lambda_{0}+
{2\Gamma(d/2)\over \pi^{d\over 2}}(1-2^{1-d})\zeta(d)
\beta^{-d}.
}
$$
$\Lambda_{0}$ is a regularized vacuum energy that will cancel after summing all
contributions.
The total result for the one-loop free energy is thus
\eqn\oneloop{
{\cal F}(\beta)_{\rm 1-loop}=-{2\Gamma(d/2)\over \pi^{d\over 2}}\zeta(d)(1-2^{-d})
({\cal D}_{\rm max}-2){N}^{2}\beta^{-d}.
}

\fig{Feynman diagrams contributing to the two-loop canonical free energy. Solid lines 
represent fermions, wavy lines
gauge bosons and dashed lines Fadeev-Popov ghosts.}{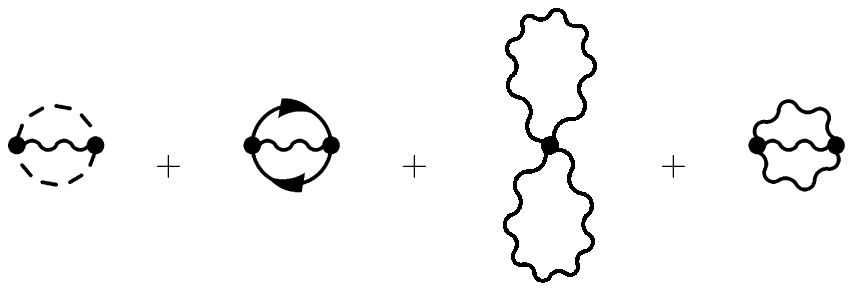}{12truein} 

Next we get the two-loop corrections to this result. Thus, we must
sum the contributions of the four Feynman diagrams of Fig. 1. corresponding 
to ${\cal N}=1$ SYM
in $\CD={\cal D}_{\rm max}$ where internal momenta is restricted to a 
$d$-dimensional space-time
(one of whose directions is the compactified Euclidean time). Proceeding 
this way and after
some algebra we find the contribution of each independent diagram (using the 
Feynman-'t Hooft gauge)
$$
\eqalign{
{\cal F}_{1}&={1\over 4}g^{2}_{YM}N^{3}\left(\sumint_{(P)}[d^{d}P]{1\over 
P^{2}}\right)^{2}, \cr
{\cal F}_{2}&={1\over 4}g^{2}_{YM}N^{3}({\cal D}_{\rm max}-2){\rm Tr\,}{\bf 1}\left\{
\left(\sumint_{\{P\}}[d^{d}P]{1\over P^{2}}\right)^{2}-2
\left(\sumint_{\{P\}}[d^{d}P]{1\over P^{2}}\right)
\left(\sumint_{(P)}[d^{d}P]{1\over P^{2}}\right)\right\}, \cr
{\cal F}_{3} &= {1\over 4} g_{YM}^{2}N^{3}{\cal D}_{\rm max}({\cal D}_{\rm max}-1)
\left(\sumint_{(P)}[d^{d}P]{1\over P^{2}}\right)^{2}, \cr
{\cal F}_{4} &= -{3\over 4} g_{YM}^{2}N^{3}({\cal D}_{\rm max}-1)
\left(\sumint_{(P)}[d^{d}P]{1\over P^{2}}\right)^{2}.
}
$$
Here ${\rm Tr\,}{\bf 1}$ is the dimension of the spinors of the maximal SYM 
theory which in all the cases under study (${\cal N}=1$ in 
${\cal D}_{\rm max}=10,6,4$) equals ${\cal D}_{\rm max}-2$.
Using this fact we can add all the above contributions and find the following expression
for the two-loop free energy density
\eqn\twoloops{
{\cal F}(\beta)_{\rm 2-loop}={1\over 4}g^{2}_{YM}N^{3}({\cal D}_{\rm max}-2)^{2}
\left(\sumint_{(P)}[d^{d}P]{1\over P^{2}}-
\sumint_{\{P\}}[d^{d}P]{1\over P^{2}}\right)^{2}.
}

The integrals appearing between brackets contain both the zero and finite 
temperature part
of ${\cal F}(\beta)_{\rm 2-loop}$. They can be easily computed by performing 
the sum first,
to give
\eqn\inttl{
\eqalign{
\sumint_{(P)}[d^{d}P]{1\over P^2}&=\int{d^{d-1}p\over (2\pi)^{d-1}}{1\over 2\omega_p}+
\int{d^{d-1}p\over (2\pi)^{d-1}}{N_{p} \over \omega_{p}}, \cr
\sumint_{\{P\}}[d^{d}P]{1\over P^2}&=\int{d^{d-1}p\over (2\pi)^{d-1}}{1\over 2\omega_p}-
\int{d^{d-1}p\over (2\pi)^{d-1}}{n_{p} \over \omega_{p}},
}
}
where $\omega_{p}=p$
and $N_{p}$, $n_{p}$ are the Bose-Einstein and Fermi-Dirac 
distribution functions respectively
$$
N_{p}={1\over e^{\beta\omega_{p}}-1}, \hskip 1cm n_{p}= {1 \over e^{\beta\omega_{p}}+1}.
$$

The first thing to be said about eq. \twoloops\ is that the zero temperature 
(ultraviolet 
divergent) contribution cancels out between the bosonic and the fermionic integral. 
This is just a consequence of supersymmetry since the vacuum energy should 
not be corrected
at zero temperature if supersymmetry is to be preserved by the vacuum \refs\witsusy. 
This is the reason why this cancellation occurs not only for the conformal $\CN=4$, 
SYM$_4$ (${\cal D}_{\rm
max}=10$) \refs\tf, but for all SYM theories under study.

Notice that although all SYM theories in dimension higher than 4 are 
non-renormalizable,
the two-loop finite temperature free energy is well behaved in the ultraviolet. 
This is a consequence of the fact that the ultraviolet region in the 
thermal integrals is effectively cut-off 
for momenta $p\gg T$ and therefore the temperature dependent part of the 
amplitudes is, to a 
great extent, insensitive to ultraviolet ambiguities. Of course, divergences 
should have been taken care of in the zero temperature sector by an appropriate cut-off 
in momenta $\Lambda$  (although some protected observables, like the vacuum energy, 
will be finite due to supersymmetry). In that case, consistency will require
$T<\Lambda$.

To get an analytical expression for the free energy we can evaluate the 
integrals appearing in 
\inttl\ for generic values of the
dimension\foot{Actually, the fermionic integral can be written in terms of 
the bosonic one using 
$$
{2\over e^{2\beta\omega_{p}}-1}={1\over e^{\beta\omega_{p}}-1}-{1\over 
e^{\beta\omega_{p}}+1}
$$
which is just a realization of the well-known relation between the one-loop
free energy of a bosonic and a fermionic quantum field, $F_{\rm fer}(\beta)=
F_{\rm bos}(\beta)-2F_{\rm bos}(2\beta)$ \refs\os.}
$$
\eqalign{
\int{d^{d-1}p\over (2\pi)^{d-1}}{N_{p}\over \omega_{p}} &=
{2^{2-d}\pi^{-{d-1\over 2}}\over \Gamma\left({d-1\over 2}\right)}
\zeta(d-2)\Gamma(d-2)\beta^{2-d}, \cr
\int{d^{d-1}p\over (2\pi)^{d-1}}{n_{p}\over \omega_{p}} &=
(1-2^{3-d}){2^{2-d}\pi^{-{d-1\over 2}}\over \Gamma\left({d-1\over 2}\right)}
\zeta(d-2)\Gamma(d-2)\beta^{2-d},
}
$$
so after substituting in \inttl\ and \twoloops\ we get
$$
{\cal F}(\beta)_{\rm 2-loop}=g^{2}_{YM}N^{3}\left[({\cal D}_{\rm max}-2){2^{2-d}
\zeta(d-2)
\over \pi^{d-1\over 2}\Gamma\left({d-1\over 2}\right)}(1-2^{2-d})
\Gamma(d-2)\right]^{2}\beta^{4-2d}
$$
which together with the one-loop contribution can be written as
\eqn\otloops{
\eqalign{
{\cal F}(\beta)&=-N^{2}\beta^{-d}\left\{
{2\Gamma(d/2)\over \pi^{d\over 2}}\zeta(d)(1-2^{-d})
({\cal D}_{\rm max}-2)\right. \cr &- \left.
g^{2}_{YM}N\left[({\cal D}_{\rm max}-2){2^{2-d}\zeta(d-2)
\over \pi^{d-1\over 2}\Gamma\left({d-1\over 2}\right)}(1-2^{2-d})
\Gamma(d-2)\right]^{2}\beta^{4-d}+{\cal O}[(g_{YM}^{2}N)^{2}]\right\}.
}
}
Incidentally, the two-loop correction to the free energy is always positive
for $d>1$, so it always tends to increase the (negative) one-loop contribution.

From \otloops\ we see explicitly how corrections to the one-loop result 
come in powers of the 't Hooft coupling $g_{YM}^{2}N$ which is kept fixed in the
large-N limit. Since in $d$-dimensions the Yang-Mills coupling constant $g_{YM}^{2}$
has dimension of (Energy)$^{4-d}$ the condition for the perturbative expansion 
to be reliable is the effective dimensionless coupling at a given temperature to 
be small
$$
g_{eff}^{2}=g^{2}_{YM}N\beta^{4-d}\ll 1.
$$
Notice that for $d>4$, for which the SYM theory is non-renormalizable, perturbative
corrections to the free energy will be governed by a small parameter in the low 
temperature limit,
corresponding to the fact that the theory is well behaved in the infrared. 
Actually, higher order corrections to formula \otloops\ have better and better
infrared behavior as we increase the order in perturbation theory. At the same
time the ultraviolet divergences worsen in the zero temperature sector, 
while in the temperature-dependent part of the amplitude the ultraviolet 
behavior is smoother due to the presence of Boltzmann factors that 
effectively cut-off momenta beyond a scale of order\foot{The better
ultraviolet behavior of the temperature-dependent sector of the theory does not 
guarantee its finiteness in higher loops; for example, the free energy in 
SYM$_{10}$ is ultraviolet divergent at three loops as can be seen by thermal 
averaging the one-loop effective action, which contains a $F^{4}$ term that 
scales quadratically with the ultraviolet cut-off. The
resulting thermal averaged divergent part is of order $\CO[(g_{YM}^2N)^{2}]$
\refs\rty, as corresponds to a three-loop contribution. I thank A. Tseytlin for 
pointing this out to me.} $T$.

From the two-loop canonical free energy we can obtain the corrections to the equation
of state of SYM$_{d}$. We first compute the canonical entropy density 
$\CS(T)$ as a function of
the temperature and invert it up to terms of order ${\cal O}[(g_{YM}N)^2]$ to get
$$
T(\CS)= \left({\CS\over N^2 \CF_{0} d}\right)^{1\over d-1}+g_{YM}^2 N\,
{ (2d-4)\CF_{1}\over
d(d-1) \CF_{0}}\,\left({\CS\over N^2 \CF_{0} d}\right)^{d-3\over d-1}+\CO[(g_{YM}^2N)^2]
$$
where the numerical coefficients $\CF_{0}$, $\CF_{1}$ are defined from the free energy
\otloops\ by $\CF=-\CF_{0}N^{2}T^{d}+(g^{2}_{YM}N) N^2 \CF_{1}T^{2d-4}$.
Now we can substitute into the internal energy density $\CE=\CF+T\CS$ with the result
$$
\CE=(d-1)\CF_{0}N^{2} \left({\CS\over N^2 \CF_{0} d}\right)^{d\over d-1}\left[
1+(g_{YM}^{2}N){\CF_{1}\over (d-1)\CF_{0}}
\left({\CS\over N^2 \CF_{0} d}\right)^{d-4\over d-1}\right]+\CO[(g_{YM}^{2}N)^{2}].
$$

\subsec{The $d=4$ case}

We will now concentrate our attention in the four-dimensional case. Taking $d=4$ in 
\otloops\ we find
\eqn\tlfd{
{\cal F}(\beta)_{d=4}=-{1\over 8}N^{2}\beta^{-4}(\CD_{\rm max}-2)
\left[{\pi^{2}\over 6}-{1\over 32}(\CD_{\rm max}-2)g_{YM}^{2}
N\right].
}
As a check, we can evaluate \otloops\ for the superconformal 
${\cal N}=4$ SYM$_4$ theory, which is obtained by dimensional reduction of
${\cal N}=1$ in SYM$_{10}$ (i.e. ${\cal D}_{\rm max}=10$) to give
$$
{\cal F}(\beta)_{{\cal N}=4}=-N^{2}\beta^{-4}\left({\pi^{2}\over 6}-{1\over 4}g_{YM}^{2}
N\right)
$$
which indeed agrees with the result of ref. \refs\tf. In the equation of state,
the loop correction just results in a renormalization of the
overall numerical factor  (this also happens in the
non-conformal cases $\CD_{\rm max}=6,4$)
$$
\CE_{\CN=4}={\pi^2\over 2}\left({3\over 2\pi^2}\right)^{4\over 3}
\left(1+{g^{2}_{YM}N\over 2\pi^2}\right)
\CS^{4\over 2}N^{-{2\over 3}} .
$$

Generically, the next contribution to \otloops\ is naively
given by three-loop diagrams of order ${\cal O}[(g_{YM}^2N)^2]$. 
However in four dimensions,
as it happens in QCD \refs\rkapnp\refs\az, at three loop level there are already 
uncanceled infrared divergences that have to be cured by summing over ring diagrams.
This gives a non-analytic (of order $\CO[(g_{YM}^2N)^{3/2}]$) contribution to the 
free energy, representing a mild failure of perturbation theory due to the infrared 
ambiguities. The evaluation of this term is essentially equivalent to dressing 
the $A_{0}^a$ and scalar propagators in loops by introducing the effect of Debye 
screening and thermal mass for the scalars. To leading order in the 't Hooft
coupling, the electric (Debye) mass can be easily computed from the static limit
of the one-loop self-energy to give
\eqn\elmass{
\eqalign{
m_{\rm el}^{2} \equiv \lim_{\vec{p}\rightarrow 0}
\Pi_{00}^{aa}(0,\vec{p}) ={1\over 4}({\cal D}_{\rm max}-2)g_{YM}^{2}NT^{2}
+\CO[(g_{YM}^{2}N)^{2}],
}
}
while for the scalars we have
\eqn\scmass{
m_{\phi}^{2} \equiv \lim_{\vec{p}\rightarrow 0} \Pi^{aa}(0,\vec{p})=
{1\over 8}(\CD_{\rm max}-2) g^2_{YM}N T^2+ \CO[(g_{YM}^{2}N)^{2}].
}

In order to get the $\CO[(g_{YM}^2N)^{3/2}]$ terms in the free energy we use
the technique of ref. \az, and rewrite the original Lagrangian density as
\eqn\lagd{
\CL_{{\rm SYM}_{4}}= \left(\CL_{{\rm SYM}_{4}}+\half m_{\rm el}^{2}{\rm Tr}A_{0}^{2}
\delta_{p_{0},0}+
\half m_{\phi}^{2}\sum_{i=1}^{n_{s}}{\rm Tr}\phi_{i}^{2}\right)-
\half m_{\rm el}^{2}{\rm Tr}A_{0}^{2}\delta_{p_{0},0}-
\half m_{\phi}^{2}\sum_{i=1}^{n_{s}}{\rm Tr}\phi_{i}^{2}
}
where $\phi_{i}$ are the $n_{s}=\CD_{\rm max}-4$ adjoint scalars in the theory
and the electric mass only affects to the zero-frequency component of the 
$A_{0}^a$ field (cf. \refs\rkapnp). The strategy now is to treat the last two 
terms as a perturbation to the Lagrangian density between brackets. This 
results in a reorganization of perturbation theory in which the 
ring-diagram contribution can be easily evaluated.

The first thing will be to compute again the one-loop free energy density, 
including now
the effect of the masses in the Lagrangian \lagd\ and, at the same time, adding new
one-loop diagrams containing vertices associated with the counterterms. Expanding
the results up to order $\CO[(g_{YM}^2N)^2]$ we find\foot{The trick of dimensional
reduction is no longer useful here because the thermal mass distinguishes 
between scalar and gauge boson propagators. Thus we have to compute all diagrams
separately.} 
\eqn\c{
\CF(\beta)_{\rm 1-loop}^{\rm resum}=\CF(\beta)_{\rm 1-loop}+
{1\over 24 \pi}
N^2 T\left[m_{\rm el}^{3}+(\CD_{\rm max}-4)m_{\phi}^{3}\right]+\CO[(g_{YM}^2N)^2]
}
with $\CF(\beta)_{\rm 1-loop}$ given by \oneloop. Proceeding similarly with the
two-loops diagrams of Fig. 1, we get
$$
\CF(\beta)_{\rm 2-loop}^{\rm resum}=\CF(\beta)_{\rm 2-loop}-
{1\over 8 \pi}
N^2 T\left[m_{\rm el}^{3}+(\CD_{\rm max}-4)m_{\phi}^{3}\right]+\CO[(g_{YM}^2N)^2].
$$
So we are left with the following final result for the ``$2\half$-loop'' 
contribution to the free energy density
\eqn\dym{
\CF(\beta)_{2\half{\rm-loop}}=-{1\over 12\pi}
N^2 T\left[m_{\rm el}^{3}+(\CD_{\rm max}-4)m_{\phi}^{3}\right]
}
where the values of the thermal masses are given by eqs. \elmass\ and \scmass.
It is important to notice that this term is always negative for all 
$4\leq\CD_{\rm max}\leq 10$. 

The only thing left now will be to add  \dym\ to the two-loop result
\tlfd. In particular, doing so for the 
superconformal $\CN=4$, SYM$_{4}$ and evaluating the numerical coefficients,
we find ($\lambda^2\equiv g^{2}_{YM}N$)
$$
\CF(\beta)_{\CN=4}=-N^2 T^4[1.645-0.250 \lambda^2 + 0.234 \lambda^3+
\CO(\lambda^4)].
$$

Next terms in the perturbative expansion in four dimensions will be of order 
$\CO(\lambda^4)$ and $\CO(\lambda^4 \log \lambda)$ and can be also evaluated 
using the strategy employed in \refs\az\ (or up to order $\CO(\lambda^5)$ using 
\refs\rfive\refs\rfived).  However, finite temperature 
perturbation theory is expected to break down at order $\CO[(g_{YM}^2N)^{3}]$
\refs\rkap. Since this failure of perturbation theory is associated with 
the infrared sector of the theory, supersymmetry is not expected to solve
the problem or even improve the situation. As in QCD \refs\rbraaten\refs\rfived,
some kind of non-perturbative analysis will be needed in order to compute 
higher orders. Actually, the general structure of the series in $\lambda$ is 
important in trying to decide whether there is a phase transition occurring at some
intermediate value of the 't Hooft coupling that precludes the extrapolation of
supergravity physics into the gauge theory domain \refs\rml.

\subsec{SYM$_{2}$ thermodynamics on ${\bf S}^{1}_{L}\times {\bf R}$}

When $d\leq 3$ the analysis of the quantum corrections to the one loop
free energy gets additional complications due to the hard infrared divergences
that afflict super-renormalizable theories. For $d=3$, we see that 
expression \otloops\ diverges
because of a $\zeta(1)$ factor. In principle this can be cured, as usual, by 
computing the thermal masses and inserting them into the propagators, thus regularizing
the low-momentum behavior of the Feynman integrals. 
However, in the three-dimensional case the computation of the electric mass 
has to be done with extra care, since the one-loop
corrections to the propagators are already infrared divergent. Thus, the electric mass
has to be evaluated  self-consistently {\it \`a la} Hartree-Fock \refs\rdhoker. 
Anyway, we will not dwell in this case any further.

The two-dimensional case, on the other hand, is more interesting from several points
of view. The one that will concern us here 
is that $\CN=8$, SYM$_{2}$ describes the world-volume dynamics
of Matrix strings \rmt, a non-perturbative definition of Type-IIA superstrings. 
Naively, \otloops\ is ill-defined for $d=2$ due to the endemic infrared divergences 
of low-dimensional field theories. There are several ways in which this divergence 
can be regularized. Here we will get rid of the problem by putting the system in 
finite box\foot{Actually, if we take the $d\rightarrow 2$ limit in expression 
\otloops\ we get a finite result with a two-loop correction independent of the 
temperature. However, since dimensional regularization is not reliable in dealing 
with infrared divergences we will not follow this procedure.} of 
length $L=2\pi R$. We will assume that the thermal wavelengths of the fundamental fields
are much smaller than the global length of the box $\beta\ll L$ and restrict our 
analysis to the sector without Wilson lines (the ``long strings'' that characterize 
the matrix string phase). 
Once this is done, the only change in the computation of Feynman diagrams
is that continuous space momentum is discretized in units of $1/R$ and the momentum 
integrals 
have to be replaced by discrete sums
\eqn\dic{
\sumint [d^2P] \longrightarrow 
 {1\over L}\sum_{n\in{\bf Z}} {1\over \beta}
\sum_{\omega_{m}}
}
where the second sum is, as usual, over integer or half-integer Matsubara 
frequencies depending
on the bosonic or fermionic character of the propagating field.

In the one-loop approximation the relevant bosonic and fermionic determinants have been 
already computed in \refs\rbvm\ and the resulting one-loop free energy 
density can be cast 
in terms of modular functions
$$
{\cal F}(\beta,L)_{\rm 1-loop}= -{1\over L \beta}N^{2}({\cal D}_{\rm max}-2)\log\left[
{\theta_{4}\left(0\left|i{L\over \beta}\right.\right)\over
\eta^{3}\left(i{L\over \beta}\right)}\right] \simsub_{\beta \ll L} 
-{\pi \over 4} (\CD_{\rm max}-2)
N^2 \beta^{-2}.
$$
In computing the bosonic determinant, and in order to keep the argument of the logarithm
dimensionless, we have added a $\beta$-independent  counterterm. In the infinite 
volume limit $L\rightarrow\infty$ we recover the one-loop result obtained in Sec. 2.2.

Let us now go to the two-loop case. To compute the contribution 
to the free energy density
we can use formula \twoloops\ provided we substitute the integration by the
sum according to \dic. After making so, we find 
\eqn\dif{
\eqalign{
\left(
\sumint_{(P)}[d^{2}P]{1\over P^2}-\sumint_{\{P\}}[d^{2}P]{1\over P^2}\right) 
&\rightarrow
{1\over L\beta}
\sum_{m,n}{}'(-1)^{n}\left[{4\pi^2 m^2\over L^2}+{4\pi^2 n^2\over 
(2\beta)^2}\right]^{-1}
\cr
& = {1\over L\beta}\, Z\left[\matrix{0 & 0\cr 0 & \half}\right](2)_{\Phi}
}
}
where we have made use of the Epstein zeta-function \refs\repstein\ and $\Phi(m,n)$ 
represents the quadratic form between the square brackets. It is interesting to notice
here that the resulting {\it regular} zeta-function arises as the difference of two
{\it singular} zeta-functions with $\zeta(1)$-like divergences which cancel out. This
is again due to the non-renormalization of the vacuum energy for supersymmetric
theories. 

Actually, the zeta-function in \dif\
can be written itself in terms of ordinary modular functions (see first article in
\refs\repstein), so at the end we can write
\eqn\tll{
\eqalign{
{\cal F}(\beta,L)_{\rm 2-loops}& 
={1\over 4\pi^{2}}g_{YM}^{2}N^{3}({\cal D}_{\rm max}-2)^{2}
\log^{2}\left[{\theta_{2}\left(0\left|i{L\over 2\beta}\right.\right) \over
\eta\left(i{L\over 2\beta}\right)}\right] \cr
& \simsub_{\beta\ll L} {N^2 \over 576} g_{YM}^{2}N (\CD_{\rm max}-2)^2
{L^2\over \beta^2}
}
}
in such a way that one arrives at the following expression for the (1+2)-loop free energy density
$$
\eqalign{
{\cal F}(\beta,L)&=-{1\over L\beta}N^{2}\left\{({\cal D}_{\rm max}-2)\log\left[
{\theta_{4}\left(0\left|i{L\over \beta}\right.\right)\over
\eta^{3}\left(i{L\over \beta}\right)}\right]\right. \cr & - \left.
{1\over 4\pi^{2}}({\cal D}_{\rm max}-2)^{2} (g_{YM}^{2}N) (L\beta)
\log^{2}\left[{\theta_{2}\left(0\left|i{L\over 2\beta}\right.\right) \over
\eta\left(i{L\over 2\beta}\right)}\right] +{\cal O}[(g_{YM}^{2}N)^2]\right\} \cr
&\simsub_{\beta \ll L} -{\pi \over 4}(\CD_{\rm max}-2)N^2\beta^{-2}
\left[1-{1\over 144\pi}
(\CD_{\rm max}-2)g^{2}_{YM}N L^{2}\right].
}
$$
According to this formula, the natural effective dimensionless coupling in the 
large $L$ limit is now $g_{eff}^{2}=(g_{YM}^{2}N) L^2 $. The analysis with be
reliable when $1\gg (g_{YM}^{2}N) L^2\gg (g_{YM}^{2}N) \beta^2$.
Again, the equation of state in the two dimensional case can be computed when 
$\beta\ll L$, with the result
\eqn\estd{
\CE={\CS^2\over \pi N^2(\CD_{\rm max}-2)}\left[1+{(\CD_{\rm max}-2)\over
144\pi }g^{2}_{YM}NL^2\right].
}

\newsec{Conclusions and outlook: Hagedorn transition from SYM thermodynamics?}

In the present paper, the thermodynamics of supersymmetric Yang-Mills theories 
with 16, 8 and 4 supercharges was studied in any dimension $d\geq 4$. We computed the
two-loop correction to the free energy for these theories and found that it 
always has opposite sign to the leading (negative) one-loop result. In the 
four-dimensional case we also evaluated the correction to the free energy 
arising from the resummation of the ring diagrams, using the technique 
of ref. \refs\az, and found it to be negative.
For lower dimensional ($d\leq 3$) SYM theories, the computation is plagued with 
infrared divergences that have to be regularized somehow. We studied in detail  
the two-dimensional case at finite volume (to regularize these infrared divergences) 
in the high temperature limit. Again we found a positive two-loop
correction which scales as $T^2$ with an effective dimensionless coupling
given by $g_{YM}^{2}NL^{2}$.

Before closing, let us make some remarks on the potential use of SYM thermodynamics
in clarifying the issue of the Hagedorn transition. On general grounds, one can 
expect two possible resolutions to the Hagedorn problem: either non-perturbative 
effects drive the critical temperature to a maximum reachable temperature for 
the system or new fundamental degrees of freedom appear at high energies, 
thus providing a picture for a phase transition (or a smooth crossover, 
depending on the details of the dynamics). Although at present there are 
no clear evidences as to which one of the two alternatives is physically 
realized in string/M-theory, some results \refs\rsathi\refs\rabkr\ and our 
still incomplete knowledge of the theory seem to hint in the direction of 
the second one.

D-instanton corrections to the thermodynamical potentials
have been studied in \refs\rbvmd\ with the result that they do not modify
the critical behavior at the Hagedorn temperature. More recently, the authors
of refs. \refs\rabkr\ have included  non-perturbative semiclassical ingredients
in the analysis of the physics of the Hagedorn transition at finite volume 
through the Horowitz-Polchinski 
correspondence principle \refs\rhp, getting a picture in which the Hagedorn 
phase is bounded at high energies by a black hole phase.  A similar 
situation occurs for a string gas on AdS backgrounds where,
in the canonical ensemble, the Hagedorn transition is ``screened'' by the
formation of an AdS black hole \refs\rbkr.

A second approach to the problem would start with a non-perturbative formulation 
of string theory in terms of M-theoretic degrees of freedom, as it has been 
proposed in \refs\rsathi.
Let us momentarily adhere ourselves to this latter  path and, starting with the 
non-perturbative definition of the Type-IIA superstring provided by Matrix 
strings \refs\rmt, study the world-volume thermodynamics of Type-IIA strings
in the microcanonical ensemble. The world-volume theory is governed by 
$\CN=8$ SYM$_{2}$ with the Yang-Mills coupling constant given by 
$g_{YM}^{2}=1/(g_{s}^2\alpha'$), with $g_{s}$ the string coupling constant. On the
other hand, free field configurations are determined by the overall scale
$\alpha^{'}$. In the infrared, $E\ll g_{YM}$,  the physics is dominated by
``long string'' excitations along the flat directions. It is in this regime in
which Matrix strings reproduce, in the large-N limit, 
the multi-string Type-IIA ensemble \refs\rmt\refs\rgrisem.
If the energy is increased, the system will begin to be excited along non-flat 
directions as well. At energies $E\gg g_{YM}$ the potential terms in the 
$\CN=8$ SYM$_{2}$ theory will behave as a small perturbation and the system 
will enter a perturbative regime. Thermodynamics there is well defined, as we have
seen from the previous analysis.

It is tempting to try to make some connection between these two world-sheet 
regimes and the low/high energy regimes in the target space string theory.
At low energies we have perturbative Type-IIA superstring theory that, in
ten open space-time dimensions, we know is characterized at high energies
by a negative specific heat phase. This negative specific heat phase is
viewed as a breakdown of equipartition in energy, in the sense that most of
the energy of the string ensemble is stored into one (or a small number) of 
highly excited strings \refs\rtherm\refs\rmicro. 
From the philosophy of M-theory it seems quite reasonable
to expect that if too much energy is stored into a single string some transition to
non-perturbative (maybe eleven-dimensional) physics should take place, 
putting an end to the negative specific phase. Alternatively, a black hole 
could be formed before the system leaves the string  regime \refs\rabkr. 
In any case, the final conclusion would be that
the Hagedorn phase will be bounded by a new phase into which the system will decay
either via a smooth crossover or a phase transition.

In the case at hand, however, it is not clear how to connect  the
world-volume theory with some kind of target picture. One of the difficulties lies
in the fact that Matrix strings are formulated in the light-cone gauge, in which 
the space-time interpretation is rather obscure. Nonetheless,  one can 
naively argue that the
negative specific heat phase at intermediate energies 
$\alpha'^{-1/2}<E\ll g_{s}^{-1}
\alpha'^{-1/2}$ is bounded at high energy $E\gg g_{s}^{-1}\alpha'^{-1/2}$ by 
a new phase with regular thermodynamics (i.e. positive specific heat) effectively 
described
by a perturbative two-dimensional $U(N)$ supersymetric Yang-Mills theory with 
sixteen supercharges in the large-N limit. If this were so, the transition between 
the low energy  string phase and the new high energy phase would be through a 
first order 
phase transition across the unstable (negative specific heat) phase (cf. 
Carlitz in \refs\rtherm). The critical points would be determined by the
Maxwell rule for the entropy, provided the complete profile of the microcanonical 
temperature $T(E)$ is known. 

The space-time interpretation of such a phase is far from being straightforward. 
In the SYM$_{2}$ perturbative regime (or directly in the free limit 
$g_{YM}^2\sim (g_{s}^2\alpha')^{-1}
\rightarrow 0$) the two-dimensional action is that of sigma-model in a 
``non-commutative''
target space with matrix coordinates $X^{\mu}\in {\bf Adj}[U(N)]$.  
Whether this indicates
that the Hagedorn transition corresponds\foot{At least in those cases in 
which it is not
preceded by the formation of black holes due to the corresponding principle. 
Actually,
we can tune the string coupling constant $g_s$, the volume and the total energy 
in such 
a way that the system avoid the correspondence line and thus we prevent the 
formation
of black holes.} to the nucleation of non-commutative bubbles in a commutative 
space-time 
is something that it is difficult to decide with our present knowledge of 
the theory. 
One of the problems to be clarified will be, for example, how the target space volume 
dependence of the extensive quantities emerges as a function of $N$. In any case, 
we stress that this extrapolation of the world-sheet 
picture to space-time physics is very
speculative, and should be tested by a detailed computation. We hope to report on this
elsewhere.

In a sense, this picture can be regarded as dual to the one proposed in 
\refs\rsathi. There, the Hagedorn transition is linked to the condensation of D0-branes
and their low-energy dynamics will be $U(N)$ super quantum mechanics with sixteen 
supercharges. Both descriptions could in principle be related by performing a T-duality
along the ninth dimension and interchanging its role with the M-theory circle.

\newsec{Acknowledgements}

It is a pleasure to thank R. Dijkgraaf, I.L. Egusquiza, R. Emparan, J.L. Ma\~nes, 
M.A.R. Osorio, M. Serone, M.A. Valle-Basagoiti, E. Verlinde, H. Verlinde and specially 
J.L.F. Barb\'on for many interesting and useful discussions. 
This work has been supported by the {\it Fundamenteel Onderzoek van the Materie} 
(FOM) Foundation and by a University of the Basque Country Grant UPV 063.310-EB187/98.

\vskip 0.2cm

\noindent
{\bf Note added}

After this paper appeared in the LANL hep-th archive, I learned directly from S.-J. Rey
of his parallel and independent work with C. Kim on SYM thermodynamics, part of which 
overlaps with the results presented here and that has later appeared in \refs\rr. 
I would like also to thank A. Nieto and A. Tseytlin for their interesting remarks 
on the first version of the article.

\listrefs

\bye